# High throughput evaluation of the effect of Mg concentration on natural ageing of Al-Cu-Li-(Mg) alloys

R. Ivanov, A. Deschamps* and F. De Geuser

Univ. Grenoble Alpes, CNRS, Grenoble INP, SIMAP, 38000 Grenoble, France.

*Correspondance e-mail: alexis.deschamps@simap.grenoble-inp.fr

**Abstract**

The effect of Mg content on the natural ageing of Al-Cu-Li-(Mg) alloys has been investigated on a compositionally graded material made by linear friction welding, allowing to probe Mg contents between 0 and 0.4 at.%. High throughput time- and space-resolved characterization of natural ageing kinetics has been achieved using small-angle X-ray scattering, supplemented by differential scanning calorimetry. Natural ageing results mainly in the formation of Cu-rich clusters whose characteristics strongly depend on the presence of Mg. Larger precipitates of 2 nm size also form to a lesser extent above an Mg concentration of 0.1 at.%.

Al-Cu-Li alloys have become a staple of the aerospace industry with their high strength to weight ratio, good toughness and corrosion resistance [1]. Research on these alloys has been focused on promoting precipitation of the $T_1$ ($Al_2CuLi$) phase through thermo-mechanical processing [2,3] and chemical alloying [4–7]. Dislocations were shown to dramatically promote the formation of high density $T_1$ precipitates by acting as heterogeneous nucleation sites [8–10]. Mg additions above 0.2 wt. % strongly influence the precipitation pathway of the alloys through the formation of precursor phases rich in Cu and Mg during ageing, subsequently promoting formation of $T_1$ phase [7,11]. Preceding the artificial ageing treatment where these precipitation processes occur, these alloys undergo clustering processes during natural ageing. Solute clustering and the associated strengthening has been well documented in most age hardening aluminum alloys [12–14] and plays a prominent role in determining the subsequent precipitation path. Starink and Wang [15] evidenced co-clustering between Cu and Mg in Al-Cu-Mg alloys and attribute it to a strong binding energy of a Cu-Mg pair. Decreus and co-workers, and Gumbmann and co-workers, report on the presence of Cu-rich clusters in Al-Cu-Li alloys with minor addition of Mg after ageing at room temperature [8,16]. In a former study, we have evidenced that adding a minor amount of Mg (0.4 at. %) to an Al-Cu-Li alloy changed dramatically the clustering behavior of Cu atoms during natural ageing [17], so that the presence of Mg seems to play a crucial role on the ability of Cu atoms to diffuse in the Al matrix and join in clusters. However, much is still unknown on this phenomenon, particularly what is the effect of the Mg concentration on the clustering kinetics of Cu, with or without a threshold content. Thus, the aim of the present contribution is to investigate the link between Mg concentration and cluster formation kinetics during natural ageing of an Al-Cu-Li alloy with a continuous variation of Mg addition.

For this purpose, we employed a high throughput approach using a compositionally graded alloy [18,19] in which the concentration of Cu and Li are kept constant and Mg varies. A diffusion couple between an Al-1.5%Cu-3.5%Li alloy and an Al-1.5%Cu-3.5%Li-0.40%Mg (all compositions in at. %) was elaborated by linear friction welding [20] to alleviate presence of an oxide layer at the diffusion interface, in the same experimental conditions as in [21]. The narrow concentration gradient present across the planar weld interface after joining was enlarged by a diffusion heat treatment at 515 °C for 14 days. The gradient was further extended via hot rolling to a distance of about 10 mm. The diffusion couple was solution treated at 500 °C for 30 min and water quenched. The concentration of Cu and Mg across the diffusion couple measured by a Cameca electron micro-probe analysis (EPMA) instrument working at 20 keV and 300 nA is shown in Figure 1.

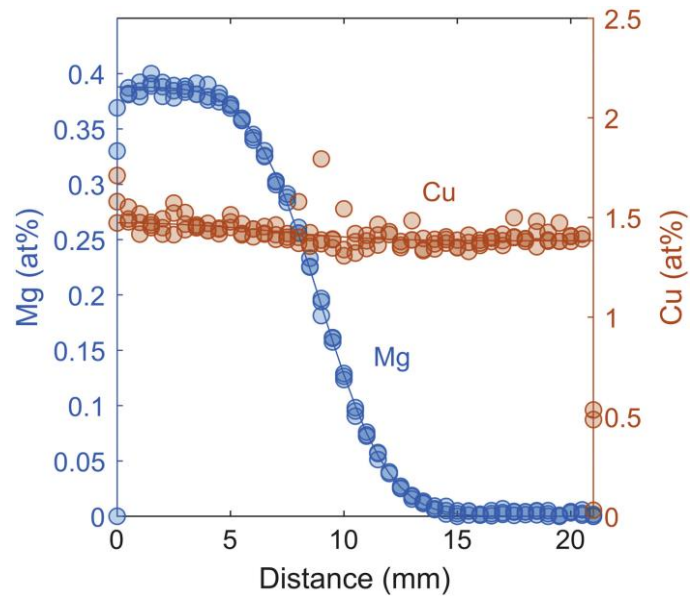

*Figure 1: Mg and Cu concentrations measured across the diffusion couple after solution treatment and quenching.*

The kinetics of clustering at room temperature was monitored in-situ during 3 days by small-angle X-ray scattering (SAXS). By displacing a 70 μm thin section of the diffusion couple continuously with respect to the X-ray beam during this monitoring, the kinetics was followed for all positions across the Mg concentration gradient with a spatial resolution of 500 μm, corresponding to a resolution in Mg chemistry of 0.02 at. % in the middle of the concentration gradient. SAXS measurements were achieved using a rotating Cu $K_\alpha$ anode with a wavelength 0.154 nm (8.048 keV). Scattering data was captured by a 2D Dectris Pilatus noiseless photon counting detector and the 2D information was radially averaged and normalized using standard procedures [22] to obtain curves of normalized intensity I(q) vs. scattering vector q. 15 points separated by 1mm were measured across the gradient, the position of which was precisely located with respect to a physical cut in the sample to correlate with the EPMA chemistry measurements. The counting time at each position was 10s during the first 10h and 100s subsequently. Variations of the cluster formation as a function of Mg concentration was further investigated by subjecting samples in the as-quenched condition to 10 °C/min heating ramps from -50°C to 350°C within a differential scanning calorimeter (DSC) using a TA Q200 instrument. Slices for DSC samples of 500 μm thickness were prepared every 1.0 mm along the Mg concentration gradient. The thickness of the DSC samples was about 1/10 of the extension of the diffusion couple; therefore, although a concentration gradient exists within a given sample, it is small in relative value, so an approximation can be made that the clustering behavior is representative of its average composition.

As shown in Figure 1, the Cu concentration across the diffusion gradient remains constant at 1.42 ±0.07 at. %. The Mg content across the diffusion gradient follows a sigmoidal curve spanning 10 mm between 0 at. % Mg on the Al-Cu-Li end and 0.39 ±0.01 at. % Mg on the Al-Cu-Li-Mg end.

The X-ray scattering factors of the diffusing element Mg (12) and of the host matrix, Al (13), are in close proximity and make scattering due to fluctuations of the Mg concentration negligible. The scattering factor of Cu (29) is significantly higher than that of the host matrix; therefore, the scattering signal observed throughout the diffusion couple predominantly arises from fluctuations in the concentration of Cu. As the Cu concentration is constant throughout the diffusion couple, the variations in SAXS intensity evidence the indirect effect of Mg on the clustering of Cu in the Al-Cu-Li alloy [6,23].

Figure 2 shows the SAXS scattering curves measured across the diffusion couple for three Mg concentrations (low, intermediate and high) after 3 days of natural ageing. Four contributions to the SAXS signal needed to be taken into account to describe the overall intensity:

$$I(q) = I_{clusters}(q) + I_{ppt}(q) + Kq^{-4} + L$$

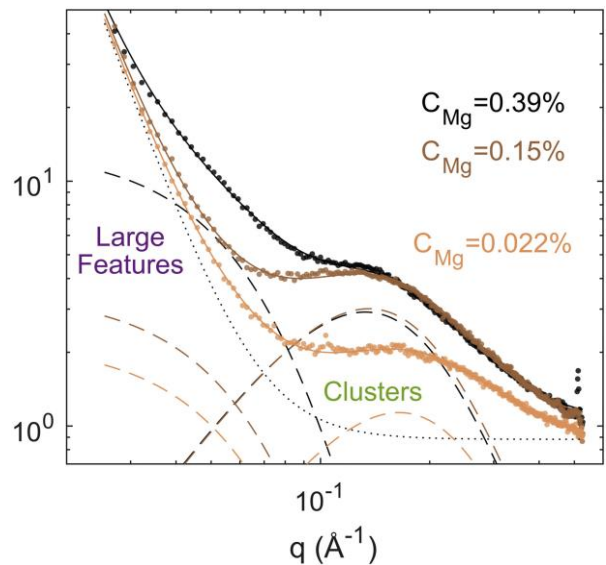

*Figure 2: Raw SAXS data (symbols) for three compositions of Mg (0.022, 0.15 and 0.39 at.%) after three days of natural ageing. The lines superimposed with the symbols show the global model fit of the data, and the interrupted lines the respective contributions calculated*

*from the fitting procedure: Porod + Laue background (dotted line), contributions from large features (dashed line, small q) and contribution from clusters (dashed line, large q).*

The last term of the equation is due to the Laue scattering from the solid solution and possible fluorescence from the residual Fe present in the sample. The $q^{-4}$ term describes the asymptotic scattering from large objects (e.g. impurities). The intensity corresponding to the decomposition of the solid solution present contributions in two distinct ranges of scattering vectors, corresponding to very small clusters for the contribution at large q, and larger objects, probably precipitates, at smaller q.

We estimated that the large objects contribution could be well described as arising from a log-normal distribution of precipitate spheres with an average radius of 2 nm and a relative dispersion of 0.2. Concerning the contribution of clusters, the classical evaluation of the SAXS intensity due to precipitates with a well-defined size, interface and resulting volume fraction [24] is not appropriate. The SAXS signal can in this case be more effectively captured using a concentration fluctuation model similar to that used by Couturier et al [25] and adapted for clusters by Ivanov et al. [17]. From this model, the quantity of clusters can be expressed as number of excess Cu atoms per unit volume. This quantity varies between approximately 10 and 50 excess Cu atoms per $nm^3$ across the Mg concentration gradient in the naturally aged state.

The evolution of the quantity and size of clusters and the quantity of precipitates (represented by mean excess Cu atoms) is presented in Figure 3. In this figure, the left column presents the evolution of these features as a function of ageing time for different Mg concentrations and the right column presents the evolution as a function of Mg concentration for different ageing times. In the as-quenched state the Mg lean and Mg rich portions of the diffusion couple have similar mean excess Cu atoms in the range of 10 $nm^{-3}$. This indicates that immediately after quench Cu rich clusters are present independently of Mg. During the first hour of natural ageing the quantity of excess Cu atoms doubles for Mg rich portions of the diffusion couple; however, the increase for the Mg lean portions in significantly lower. The majority of Cu clustering occurs during the first 10h leading to excess Cu atoms >40 $nm^{-3}$ in the Mg rich portion compared to <20 $nm^{-3}$ in the Mg lean portion. The effect of Mg concentration on the cluster size is similar to that of the cluster quantity, namely the cluster size increases together with the quantity of Cu involved in clustering. At the end of natural ageing, clusters in the Mg-lean regions have a size of approximately 0.5 nm, and in the Mg-rich region of more than 0.6 nm. The right column of Figure shows for many different ageing times the effect of Mg concentration on the characteristics of the forming clusters. These graphs clearly show that (i) there is almost no measurable lower limit for the effect of Mg on Cu clustering, if there is one it should be lower than 0.02 at. %; (ii) the effect of Mg on Cu clustering is monotonous in the investigated range and (iii) this effect seems to saturate above approximately 0.2 at. %.

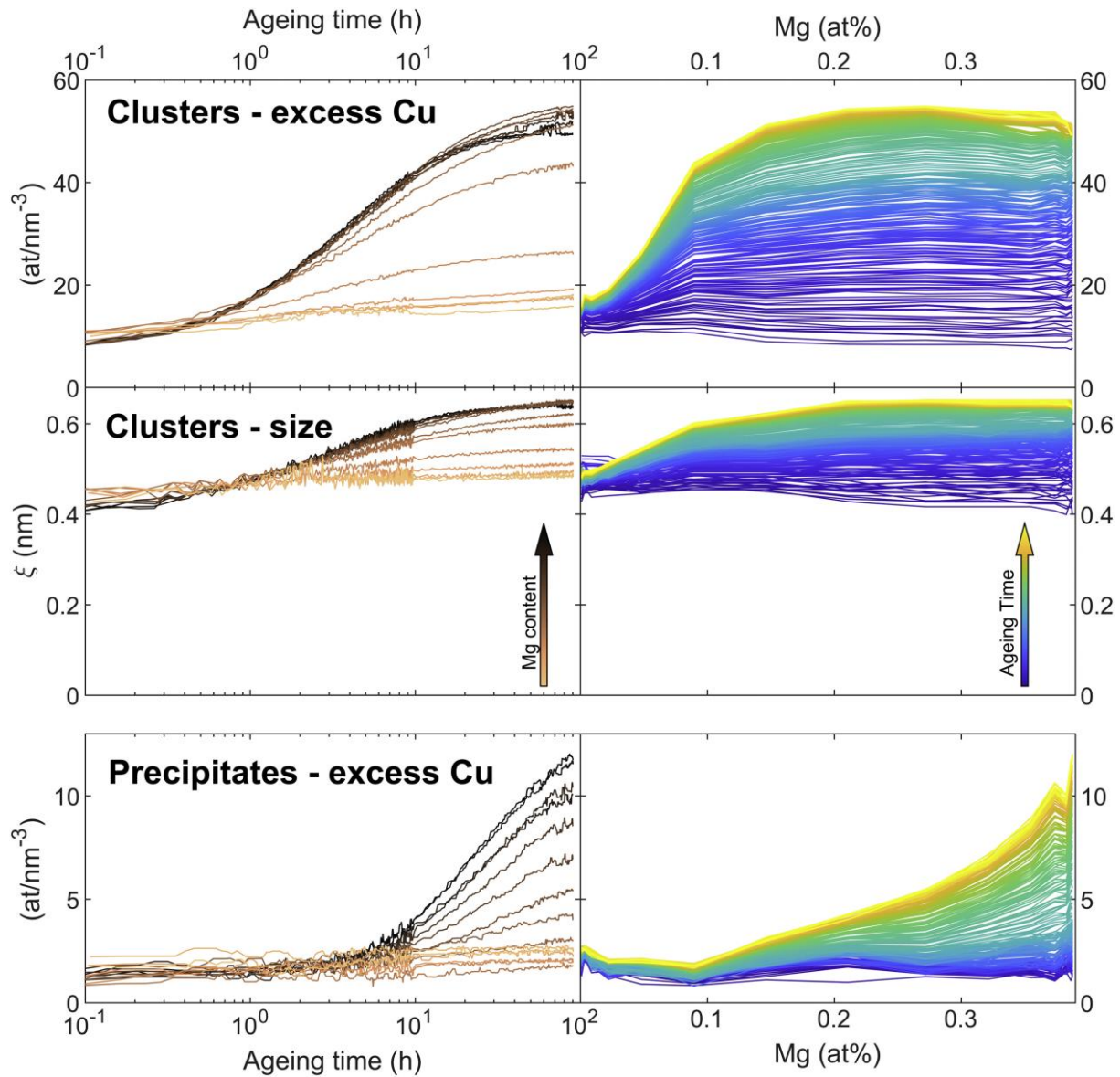

*Figure 3: Evolution of the cluster and precipitate characteristics: top figures, excess Cu atoms involved in clusters; middle figures, cluster size; bottom figures, excess Cu atoms involved in precipitates. Left column: curves for increasing Mg content as a function of natural ageing time; right column: curves for increasing ageing time as a function of Mg content.*

The behavior of the large-scale objects, that we will now name "precipitates" as a function of Mg content and ageing time is qualitatively similar but quantitatively very different. Although it corresponds to a smaller amount of excess Cu atoms as compared to the clusters, it would still represent a precipitate volume fraction of about 0.15 % at maximum if one considers a precipitate composition in Cu of 25 at. % (as in the S phase $Al_2CuMg$). These precipitates appear both later during ageing (only significantly after 10h of ageing) and as a function of Mg content (only above 0.1 at. %). This lower limit equals precisely that found by Gumbmann and co-workers [21] for the effect of Mg on precipitation of $T_1$, which they attributed to the minimum Mg content necessary to precipitate the Mg-Cu rich precursor phase at the dislocations during early ageing. Therefore, it is likely that the precipitates identified in the present study belong to the precursor phase observed by atomic STEM-HAADF imaging in [7]. Given their size, much larger relative to that of clusters, it is expected that they would take longer time at room temperature to form.

The clusters observed in the naturally aged material have formed during the decomposition of the solid solution from the as-quenched state, where the solutes are in solid solution alongside with excess vacancies. Spatially resolved DSC experiments were carried out in the as-quenched state to investigate further the relationship between Mg concentration and the clustering reaction. In these experiments, formation of clusters results in excess heat giving rise to exothermic peaks in the thermograms. The continuous heating leads to subsequent dissolution of clusters and corresponding endothermic peaks, and may lead to precipitation at higher temperatures. Figure 4 shows the thermograms across the diffusion couple in the as quenched state. They evolve with three distinct features as Mg concentration is increased. First, an exothermic peak develops at temperatures between 50 and 120 °C. The peak becomes evident at 0.09 at. % Mg and it continues to increase in magnitude with increasing Mg. This peak represents the formation of clusters as more are forming during the heating ramp with higher concentration of Mg. The second feature is a dissolution region (endothermic peak) at temperatures from 125 °C up to about 250 °C. Higher Mg concentrations cause the dissolution events to occupy a wider temperature range. Additions of Mg from 0.09 up to 0.18 at. % result in a single dissolution peak at 175 °C with larger magnitude for higher Mg. Above 0.21 at. % Mg the endothermic region shows a second peak at 215 °C, which is convoluted within the first. The magnitude of the peak at 215 °C increases when more than 0.21 at. % Mg is present in the sample. The third and final feature is the large exothermic peak at temperatures above 250 °C representing precipitation in the sample. The position of the precipitation peak shifts to lower temperatures when Mg is added to Al-Cu-Li with a notable step at around 0.18 at. % Mg. The precipitation peak corresponds to the formation of the θ' and $T_1$ phases, whose competitive formation has a strong dependence on the concentration of Mg as previously reported [7]. The 0.18 at.% transition between the broad high temperature precipitation peak (corresponding mostly to θ' formation) and the sharp low temperature precipitation peak (corresponding mostly to $T_1$ formation) corresponds well to the minimum value determined previously by SAXS kinetics and TEM observations [16], and is relatively close to the minimum value for the formation of the "large" precipitates at room temperature found in the present study.

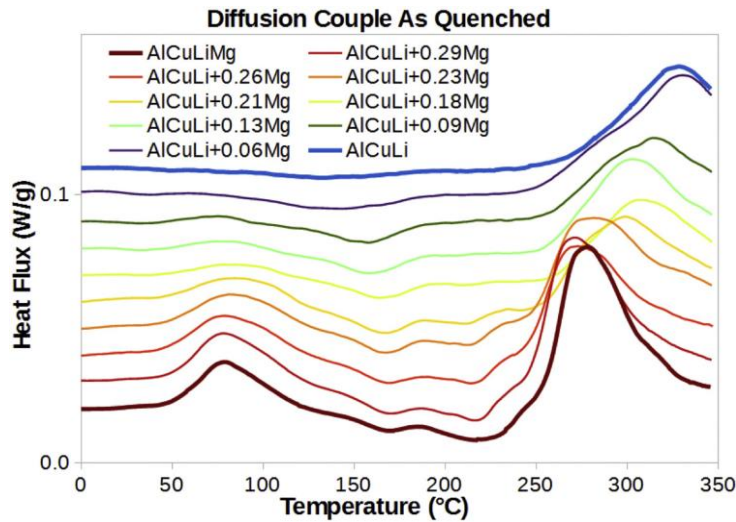

*Figure 4: DSC thermograms across the diffusion couple in the as-quenched condition*

*(exothermic effects positive)*

In summary, the influence of Mg concentration on natural ageing of Al-1.5Cu-3.5Li-xMg at. % alloy has been evaluated using a Mg diffusion couple where x varies between 0 and 0.4 at. %. Analysis performed using time and space resolved small-angle X-ray scattering and differential scanning calorimetry (DSC), reveal strong changes to clustering behaviour across the concentration gradient, as well as changes in the formation of precipitates with size ~2nm. The triggering effect of Mg addition on the clustering kinetics of Cu is observed be monotonously increasing with Mg content without a measurable minimum value, and saturates above ~0.2 at. % Mg; the formation of the precipitates occurs only above ~0.1 at. % Mg. The results on clustering show that Cu diffusion takes place faster and to a further extend with increasing Mg concentration strongly suggesting that Mg leads to more (and/or more mobile) vacancies to be available for diffusion.

## Acknowledgements

The authors would like to thank Dr Christophe Sigli for providing the alloys and fruitful discussions. Dr. Eva Gumbmann is thanked for providing the linear friction welded joint. The authors would also like to thank Dr. Florence Robaut for help acquiring the EPMA data. The authors acknowledge the financial support from Constellium C-Tec.

**Figure captions**

*Figure 1: Mg and Cu concentrations measured across the diffusion couple after solution treatment and quenching*

*Figure 2: Raw SAXS data (symbols) for three compositions of Mg (0.022, 0.15 and 0.39 at. %) after three days of natural ageing. The lines superimposed with the symbols show the global model fit of the data, and the interrupted lines the respective contributions calculated from the fitting procedure: Porod + Laue background (dotted line), contributions from large features (dashed line, small q) and contribution from clusters (dashed line, large q).*

*Figure 3: Evolution of the cluster and precipitate characteristics: top figures, excess Cu atoms involved in clusters ; middle figures, cluster size ; bottom figures, excess Cu atoms involved in precipitates. Left column: curves for increasing Mg content as a function of natural ageing time; right column: curves for increasing aging time as a function of Mg content.*

*Figure 4: DSC thermograms across the diffusion couple in the as-quenched condition (exothermic effects positive).*